\documentclass[12pt]{article}
\usepackage{amsthm,amssymb,amsmath}
\usepackage[english]{babel}

 1 
1

\newcommand{\bB}{\mathcal{B}}
\newcommand{\bL}{\mathcal{L}}
\newcommand{\bM}{\mathcal{M}}

\newcommand{\bt}{{\boldsymbol{t}}}
\newcommand{\bu}{{\boldsymbol{u}}}
\renewcommand{\d}{\operatorname{d}}
\newtheorem{teh}{Theorem}

\newcommand{\be}{\begin{equation}}
\newcommand{\ee}{\end{equation}}
\begin{document}
\title{\sc Additional symmetries and solutions of  the dispersionless KP
hierarchy\thanks{Partially supported by CICYT proyecto
BFM2002--01607 }}
\author{ Luis Mart\'{\i}nez Alonso$^{\dag}$,  Manuel Ma\~{n}as$^{\ddag}$
 \\
\emph{ Departamento de F\'{\i}sica Te\'{o}rica II, Universidad
Complutense}\\ \emph{E28040 Madrid, Spain} \\
\texttt{$^\dag$luism@fis.ucm.es}\\
\texttt{$^\ddag$manuel@darboux.fis.ucm.es}}
\date{} \maketitle
\begin{abstract}
The dispersionless KP hierarchy is considered from the point of
view of the twistor formalism.  A set of explicit additional
symmetries is characterized and its action on the solutions of the
twistor equations is studied. A method for dealing with the
twistor equations by taking advantage of hodograph type equations
is proposed. This method is applied for determining the orbits of
solutions satisfying reduction constraints of  Gelfand--Dikii type
under the action of additional symmetries.
\end{abstract}

\vspace*{.5cm}

\begin{center}\begin{minipage}{12cm}

\end{minipage}
\end{center}
\newpage

\section{Introduction}

The so-called dispersionless hierarchies \cite{1}-\cite{9} provide
an interesting type of non-linear integrable models which can not
be studied by the standard schemes of the  KP theory and   require
an entirely  new approach. From the point of view of the Lax
formalism, dispersionless hierarchies arise as the quasiclassical
limits of  Lax pair equations performed by replacing operators by
phase space functions and commutators by Poisson brackets. In this
way, when dealing with dispersionless hierarchies, instead of the
associated auxiliary linear system of the standard formalism of
integrable systems the underlying equations to be solved  are of
Hamilton--Jacobi type.

 Several methods of solution of  dispersionless hierarchies
have been formulated. In \cite{3}-\cite{4} (see also
\cite{11}-\cite{11a})  Kodama and Gibbons gave a direct method
based on the use of reductions  in which the dependent variables
depend on a finite number of unknown functions. The corresponding
reduced hierarchy becomes an infinite set of compatible
hydrodynamic systems solvable by hodographic techniques.  A
$\bar{\partial}$ scheme  has been proposed by Konopelchenko et al
in \cite{12}-\cite{14}, which introduces  an associated
$\bar{\partial}$ equation of Hamilton--Jacobi type. In this paper
we deal with the twistorial method of Takasaki and Takebe
\cite{9}-\cite{10}. Two important advantages of this method are
\begin{description}
\item[1)] It provides a convenient scheme for describing the
symmetries. \item[2)] All local solutions can be determined by
means of the twistor method.
\end{description}

The main aim of this paper is to present a technique for deriving
explicit examples of both additional symmetries and solutions of
dispersionless hierarchies within the framework of the twistor
formalism. It requieres a new formulation of the twistor equations
which involves a certain type of generating functions for
canonical transformations of twistor data as well as the use of
hodograph equations. To show our strategy, we concentrate on the
dispersionless KP (dKP) hierarchy, which is the prototype of this
kind of integrable hierarchies.  Its Lax pair formulation involves
a phase space with a canonical pair of coordinates $(p,x)$ and an
associated Poisson bracket
\[
\{F,G\}=\frac{\partial F}{\partial p}\frac{\partial G}{\partial
x}-\frac{\partial F}{\partial x}\frac{\partial G}{\partial p}.
\]
It is useful to introduce an enlarged Lax formalism with a pair of
canonically conjugate variables $\bL=\bL(p,\bt)$ and
$\bM=\bM(p,\bt)$ (i.e. $\{\bL,\bM\}=1$) depending on $p$ and an
infinite set of time parameters
\[
\bt:=(t_1=x,t_2,\ldots,t_n,\ldots),
\]
which are assumed to admit expansions of the form
\begin{equation}\label{1}
\bL=p+\sum_{n\geq 1}\frac{u_n(\bt)}{p^n},\quad
 \bM=\sum_{n\geq 2}nt_n\bL^{n-1}+x+\sum_{n\geq
1}\frac{v_n(\bt)}{\bL^{n+1}},
\end{equation}
as $p\rightarrow\infty$ and $\bL\rightarrow\infty$, respectively.
The Lax equations of the dKP hierarchy are
\begin{equation}\label{2}
\frac{\partial \bL}{\partial t_n}=\{\bB_n,\bL \},\quad
\frac{\partial \bM}{\partial t_n}=\{\bB_n,\bM \},\quad n\geq 2,
\end{equation}
where
\[
\bB_n:=(\bL^n)_{\geq 0}.
\]
Here $(\mathcal{F})_{\geq 0}$ denotes the projection of a Laurent
series  $\mathcal{F}$ in the variable $p$ on the subspace
generated by the non-negative powers of $p$ (we will also use the
notation $(\mathcal{F})_{\leq -1}:=\mathcal{F}-
(\mathcal{F})_{\geq 0}$). The system of compatibility equations
\begin{equation}\label{3}
\frac{\partial \bB_m}{\partial t_n}-\frac{\partial \bB_n}{\partial
t_m}+\{\bB_m, \bB_n \}=0,\quad m\neq n,
\end{equation}
yields an infinite set of nonlinear equations for the coefficients
$u_n$ of the expansion \eqref{1} of $\bL$. In particular for
$(n,m)=(2,3)$ one gets the dKP equation (Zabolotskaya-Khokhlov
equation)
\begin{equation}\label{4}
(u_t-3uu_x)_x=\frac{3}{4}u_{yy},\quad u:=u_1,\; t:=t_3,\; y:=t_2.
\end{equation}
This is an interesting nonlinear model with   applications, in the
study of quasi-plane sound beams \cite{16a}, quasi-transonic flows
past thin wings \cite{16b} or Einstein-Weyl spaces \cite{16c}.

In the next Section we first describe in brief the twistor
approach to the solutions and symmetries of the dKP hierarchy.
Then we present a class of additional symmetries depending on
arbitrary functions of one variable, the action of which  can be
explicitly determined.  As a particular case they include the
symmetries of the dKP equation found by Dunajski, Mason and Tod in
\cite{16c}. The first part of Section 3 is devoted to a new
formulation of twistor equations which is appropriate for dealing
with the transformation laws of solutions under the action of
symmetries. In the second part of Section 3 we show how solutions
of the dKP hierarchy satisfying reduction constraints of
Gelfand-Dikii type transform under the class of additional
symmetries introduced in Section 2. Finally, some explicit
examples are worked out.

\section{Symmetries in the twistor formalism}
\subsection{Twistorial structure of the dKP hierarchy}

The twistor formalism of the dKP hierarchy is based on the
degenerate symplectic form \cite{9}
\begin{equation}\label{5}
\omega:=\d p\wedge \d x+\sum_{n\geq 2}\d  \bB_n\wedge \d t_n.
\end{equation}
which plays the role of the Gindikin bundle \cite{15} of curved
twistor theory. The form $\omega$ encodes both the Lax equations
and their compatibility conditions into the simple system
\begin{equation}\label{6}
\begin{cases} \omega=\d\bL\wedge \d\bM,\\
\omega\wedge\omega=0.
\end{cases}
\end{equation}
From the first equation we have that
\[
\d\Big(\bM \d\bL+p\d x+\sum_{n\geq 2}\bB_n \d t_n\Big)=0,
\]
so that there exists a generating function $S=S(\bL,\bt)$ for the
canonical transformation $(p,x)\mapsto (\bL,\bM)$ satisfying
\[
\d S=\bM \d\bL+p\d x+\sum_{n\geq 2}\bB_n \d t_n,
\]
or equivalently
\begin{equation}\label{7}
\bM=\frac{\partial S}{\partial \bL},\quad p=\frac{\partial
S}{\partial x},\quad \bB_n=\frac{\partial S}{\partial t_n},\quad
n\geq 2.
\end{equation}
Notice that from \eqref{1} and the first equation of \eqref{7} it
follows that $S$ can be defined as
\[
S(\bL,\bt)=\sum_{n\geq 1}t_n\bL^n-\sum_{n\geq
1}\frac{v_n(\bt)}{n}\bL^{-n}.
\]

The twistor scheme for solving the dKP hierarchy is based on the
following result \cite{9}
\begin{teh} Let $(P(p,x),X(p,x))$ be a pair of canonically
conjugate variables (i.e. \{P,X\}=1). Then

\noindent 1) Given two functions $(\bL(p,\bt),\bM(p,\bt))$ of the
form \eqref{1} such that the composite functions
$(P(\bL,\bB),X(\bL,\bB))$ have Laurent series expansions in $p$
satisfying the \emph{twistor equations}
\begin{equation}\label{8}
(P(\bL,\bM))_{\leq -1}=0,\quad (X(\bL,\bM))_{\leq -1}=0,
\end{equation}
then  $(\bL,\bM)$ gives a solution of the dKP hierarchy \eqref{2}.
The pair
\[
(P(p,x),X(p,x))
\]
is called \emph{the twistor data} of the solution $(\bL,\bM)$.

\noindent 2) Each solution of the dKP hierarchy admits a set
$(P(p,x),X(p,x))$ of twistor data.
\end{teh}

In general, we can not assume the existence of appropriate
solutions $(\bL,\bM)$ of \eqref{8}. For example, the canonical
variables
\begin{equation}\label{9}
P:=p^2x,\quad X:=\frac{1}{p},
\end{equation}
determine the  twistor equations
\[
(\bL^2\bM)_{\leq -1}=0,\quad \Big(\frac{1}{\bL}\Big)_{\leq -1}=0.
\]
which obviously have no solutions satisfying \eqref{1}.

\subsection{Symmetry transformations}

One the main features of the twistor equations is that
 the symmetry properties of the
dKP hierarchy can be formulated in a convenient way \cite{9}.
Indeed, the natural group acting on the set of twistor data
$(P(p,x),X(p,x))$ is the group of canonical transformations
generated by one-parameter groups of the form
\begin{equation}\label{10}
\begin{gathered}
\exp(s\{F,\,\cdot\}):(P,X)\mapsto (P(s),X(s)),\quad F=F(p,x),\\
P(s):=P(\exp(s\{F,\,\cdot\})p,\exp(s\{F,\,\cdot\})x),\\
X(s):=X(\exp(s\{F,\,\cdot\})p,\exp(s\{F,\,\cdot\})x),
\end{gathered}
\end{equation}
where
\[
\exp(s\{F,\,\cdot\})G:=G+s\{F,G\}+\frac{s^2}{2}\{F,\{F,G\}\}+\cdots.
\]
 It can be proved that \cite{9}
\begin{teh}
A one-parameter group of canonical transformations \eqref{10}
induces an action $(\bL,\bM)\mapsto (\bL(s),\bM(s))$ on the set of
solutions of the dKP hierarchy determined by the flow
\begin{equation}\label{11}
\frac{\partial \bL}{\partial s}=\{\bL,F(\bL,\bM)_{\leq -1}\},\quad
\frac{\partial \bM}{\partial s}=\{\bM,F(\bL,\bM)_{\leq -1}\}.
\end{equation}

\end{teh}

\vspace{0.4truecm}

Let us  consider symmetries of the dKP hierarchy generated by
double series  of the form
\begin{equation}\label{12}
F(\bL,\bM)=\sum_{i=-\infty}^{\infty}\sum_{j=-\infty}^{\infty}c_{ij}\bL^i
\bM^j.
\end{equation}
We will concentrate on the $(r+1)$-th \emph{truncated dKP
hierarchies} defined as the sets of the  first $r+1$ flows of the
dKP hierarchy ($r\geq 2$).  Thus in order to analyze their
symmetries we  may set $t_n=0,\forall n\geq r+1$, and so we may
write
\begin{equation}\label{13}
\bM=(r+1)t_{r+1}\bL^r+rt_r\bL^{r-1}+\cdots+x+\mathcal{O}\Big(\frac{1}{\bL^2}\Big).
\end{equation}
By substituting this expansion in \eqref{12},  a series expansion
of $F$ in powers of $\bL$ is obtained. Let us now investigate
those symmetries of the \newline $(r+1)$-th truncated  dKP
hierarchy which do not involve the action of higher dKP flows. To
this end, we have to avoid terms of the form $\{(\bL^n)_{\geq
0},\bL\}$
 with $n>(r+1)$ in the right-hand side of the first equation in
\eqref{11}. Hence we impose $c_{ij}=0$ for $(i+jr)>(r+1)$, so that
$F$ can be expressed in the form
\begin{equation}\label{14}
F(\bL,\bM)=\sum_{n\leq r+1}\alpha_n\Big(
\frac{\bM}{(r+1)\bL^r}\Big)\bL^n,
\end{equation}
with $\alpha_n(t)$ being arbitrary smooth functions. Furthermore,
Eq.\eqref{11} for $\bL$ can be written as
\begin{equation}\label{15}
\frac{\partial \bL}{\partial s}=\frac{\partial F}{\partial \bM}
+\{F(\bL,\bM)_{\geq 0},\bL \},
\end{equation}
and it is easy to see that only those terms in \eqref{14} with
$n\geq 1$ contribute to $\partial u /\partial s$.

Therefore, we conclude that the symmetries of the $(r+1)$-th
truncated dKP hierarchy which do not involve higher dKP flows and
define a non-trivial action on the coefficient $u$ are of the form
\begin{equation}\label{16}
F(\bL,\bM)=\sum_{n=1}^{r+1}\alpha_n\Big(
\frac{\bM}{(r+1)\bL^r}\Big)\bL^n.
\end{equation}
This means that, under these conditions,  there are essentially
$r+1$ types of symmetry generators of the $(r+1)$-th truncated dKP
hierarchy given by
\begin{equation}\label{17}
F_i(\bL,\bM):=\alpha\Big( \frac{\bM}{(r+1)\bL^r}\Big)\bL^i,\quad
i=1,\ldots,r+1,
\end{equation}
with $\alpha=\alpha(t)$ being an arbitrary function.

 The action of the one-parameter groups generated by $F_i$ on
the coefficient $u$ can be explicitly found.  Indeed, by
identifying  the coefficients of $1/p$ in both members of
\eqref{15} one gets a first-order \emph{linear} partial
differential equation for
\[
u(s,\bt):=\exp(s\{F_i,\,\cdot\})u(\bt),
\]
the integration of which provides the symmetry transformation
\[
u=u(\bt)\mapsto \tilde{u}=u(s,\bt).
\]

Let us illustrate these facts by considering the case $r=2$. We
observe that \eqref{13} implies that near points $t$ in the region
of analyticity of $\alpha$
\begin{equation}\label{18}
\alpha\Big(
\frac{\bM}{3\bL^2}\Big)=\alpha(t)+\frac{2}{3}y\alpha'(t)\frac{1}{\bL}
+\Big(\frac{1}{3}x\alpha'(t)+\frac{2}{9}y^2\alpha''(t)\Big)\frac{1}{\bL^2}
+\mathcal{O}\Big(\frac{1}{\bL^3}\Big).
\end{equation}
One finds the following results for the corresponding three
generators \eqref{17}:

\vspace{0.2 truecm}

\noindent {\bf 1. $F_1$}

\vspace{0.2 truecm}

From \eqref{15} we have
\[
\frac{\partial \bL}{\partial
s}=\alpha'\Big(\frac{\bM}{3\bL^2}\Big)
\frac{1}{3\bL}+\alpha(t)\frac{\partial \bL}{\partial x},
\]
so that
\begin{equation}\label{19}
\frac{\partial u}{\partial s}=\alpha(t)\frac{\partial u}{\partial
x}+\frac{1}{3}\alpha'(t).
\end{equation}
The solution of this equation is
\[
u=U(x+s\alpha(t),y,t)+\frac{1}{3}s\alpha'(t),
\]
where $U$ is an arbitrary function. It leads to the symmetry
\begin{equation}\label{20}
\widetilde{u}=u(x+s\alpha(t),y,t)+\frac{1}{3}s\alpha'(t).
\end{equation}

\vspace{0.2 truecm} \noindent {\bf 2. $F_2$}

\vspace{0.2 truecm} In this case \eqref{15} becomes
\[
\frac{\partial \bL}{\partial
s}=\frac{1}{3}\alpha'\Big(\frac{\bM}{3\bL^2}\Big)
+\frac{2}{3}y\alpha'(t) \frac{\partial \bL}{\partial
x}+\alpha(t)\frac{\partial \bL}{\partial y},
\]
and the equation for $u$ is
\begin{equation}\label{21}
\frac{\partial u}{\partial s}=\frac{2}{3}y\alpha'(t)\frac{\partial
u}{\partial x}+\alpha(t)\frac{\partial u}{\partial
y}+\frac{2}{9}y\alpha''(t),
\end{equation}
which has the solution
\begin{align*}
u=&U\Big(x+\frac{2}{3}sy\alpha'(t)+\frac{1}{3}s^2\alpha(t)\alpha'(t)
,y+s\alpha(t),t\Big)\\
+&\frac{2}{9}sy\alpha''(t)+\frac{1}{9}s^2\alpha(t)\alpha''(t),
\end{align*}
where $U$ is an arbitrary function. The corresponding symmetry
transformation of the dKP equation is
\begin{equation}\label{22}
\begin{aligned}
\widetilde{u}=&u(x+\frac{2}{3}sy\alpha'(t)+\frac{1}{3}s^2\alpha(t)\alpha'(t),y+s\alpha(t),t)\\
+&\frac{2}{9}sy\alpha''(t)+\frac{1}{9}s^2\alpha(t)\alpha''(t).
\end{aligned}
\end{equation}

\vspace{0.2 truecm} \noindent {\bf 2. $F_3$}

\vspace{0.2 truecm} Now Eq.\eqref{15} takes the form
\begin{align*}
\frac{\partial \bL}{\partial s}&
=\frac{1}{3}\alpha'\Big(\frac{\bM}{3\bL^2}\Big)\bL
+\Big(\frac{1}{3}x\alpha'(t) +\frac{2}{9}y^2\alpha''(t)\Big)
\frac{\partial \bL}{\partial x}\\
&+\frac{2}{3}y\alpha'(t)\frac{\partial \bL}{\partial y}+
\alpha(t)\frac{\partial \bL}{\partial t},
\end{align*}
which implies
\begin{equation}\label{23}
\begin{aligned}
\frac{\partial u}{\partial s}& =\Big(\frac{1}{3}x\alpha'(t)
+\frac{2}{9}y^2\alpha''(t)\Big) \frac{\partial u}{\partial
x}+\frac{2}{3}y\alpha'(t)\frac{\partial u}{\partial y}
+\alpha(t)\frac{\partial u}{\partial t}\\&+\frac{1}{3}\alpha'(t)u
+\frac{1}{9}x\alpha''(t)+\frac{2}{27}y^2\alpha'''(t).
\end{aligned}
\end{equation}
The solution of this equation is
\begin{align*}
u=&(c'(t))^{\frac{2}{3}}U\Big(x(c'(t))^{\frac{1}{3}}+\frac{2}{9}y^2\frac{c''(t)}
{(c'(t))^{\frac{2}{3}}}
,y(c'(t))^{\frac{2}{3}},s+c(t)\Big)\\
+&\frac{1}{9}x\frac{c''(t)}{c'(t)}+\frac{2}{27}y^2\Big(\frac{c'''(t)}{c'(t)}-
\frac{4}{3}\Big(\frac{c''(t)}{c'(t)}\Big)^2\Big),
\end{align*}
where $U$ is an arbitrary function and $c(t)$ is such that
$c'(t)=1/\alpha(t)$. Hence, by defining $T:=T(s,t)$ through the
implicit relation
\[
c(T)=s+c(t),
\]
and by taking into account that
\[
T':=\frac{\partial T}{\partial t}=\frac{c'(t)}{c'(T)},
\]
one finds that the symmetry transformation determined by
\eqref{23} is
\begin{equation}\label{24}
\begin{aligned}
\tilde{u}=&(T')^{\frac{2}{3}}u\Big(x(T')^{\frac{1}{3}}+\frac{2}{9}y^2\frac{T''}
{(T')^{\frac{2}{3}}}
,y(T')^{\frac{2}{3}},T\Big)\\
+&\frac{1}{9}x\frac{T''}{T'}+\frac{2}{27}y^2\Big(\frac{T'''}{T'}-
\frac{4}{3}\Big(\frac{T''}{T'}\Big)^2\Big)
\end{aligned}
\end{equation}

The three symmetries \eqref{20},\eqref{22} and \eqref{24} coincide
with the symmetries of the dKP equation found by Dunajski, Mason
and Tod \cite{16c} by analyzing equivalence transformations of
Einstein--Weyl spaces.

\vspace{0.2cm}

\noindent {\bf Transformation law of twistor data}

\vspace{0.2cm}

According to \eqref{10},  the dKP symmetry generated by \eqref{17}
corresponds to a canonical transformation law of the twistor data
determined by the Hamiltonian system
\begin{equation}\label{24a}
\frac{\d p}{\d s}=\{\alpha(\rho)p^i,p\},\quad \frac{\d x}{\d
s}=\{\alpha(\rho)p^i,x\},
\end{equation}
where we are denoting
\[
\rho:=\frac{x}{(r+1)p^r}.
\]
In terms of $(p,\rho)$ this system becomes
\begin{equation}\label{24aa}
\frac{\d p}{\d s}=-\frac{\alpha'(\rho)}{r+1}p^{i-r},\quad \frac{\d
\rho}{\d s}=i\frac{\alpha(\rho)}{r+1}p^{i-r-1},
\end{equation}
and by taking into account that the Hamiltonian function
\[
h:=\alpha(\rho)p^i
\]
is a constant of the motion it follows that the solution of
\eqref{24a} can be written as
\begin{equation}\label{25a}
p(s)=\displaystyle\frac{p}{(j_{\rho})^{\frac{1}{r+1}}},\quad
x(s)=(r+1)j\,p(s)^r.
\end{equation}
Here $j=j(s,\rho,h)$ is  the evolution law of the variable $\rho$.
That is to say, it is the solution of the initial value problem
\begin{equation}\label{25b}
\frac{\partial j}{\partial s}=\beta(\rho,h),\quad
j(0,\rho,h)=\rho,
\end{equation}
where
\[
\beta(\rho,h):=\frac{i}{r+1}\Big(\frac{\alpha(\rho)}{h}\Big)^
{\frac{r+1}{i}}h.
\]

 The expressions \eqref{25a} define the action of the additional symmetries \eqref{17} on the twistor
data. It is important to observe that the solution of \eqref{25b}
satisfies
\[
s=\int_{\rho}^{j(s,\rho,h)}\frac{\d \rho}{\beta(\rho,h)},
\]
and, as a consequence, one deduces that the first-order
derivatives of $j$ with respect to $\rho$ and $h$ are
\begin{equation}\label{26a}
\begin{aligned}
j_{\rho}&=\Big(\frac{\alpha(j)}{\alpha(\rho)}
\Big)^{\frac{r+1}{i}},\\
j_h&=s\Big(\frac{i}{r+1}-1\Big) \Big(\frac{\alpha(j)}{h}
\Big)^{\frac{r+1}{i}}=\Big(\frac{i}{r+1}-1\Big)\frac{s}{p(s)^{r+1}}.
\end{aligned}
\end{equation}
As we will see below,  these relations will be useful for
determining the action of the additional symmetries on the
solutions of the twistor equations.

\section{Solutions of the dKP hierarchy}

\subsection{Generating functions and hodograph equations}

We are going to present a scheme for solving twistor equations
which is particularly suitable to investigate the action of the
additional symmetries introduced in the above section. An
ingredient of our analysis is the use of a type of generating
functions for canonical transformations of twistor data \cite{17},
which allows us to introduce  hodograph type equations to
formulate part of the constraints imposed by the twistor
equations.

Let $(P(p,x),X(p,x))$ be a pair of canonically conjugate
variables, then for each positive integer $r$ we have
\[
\d P\wedge \d X=\d p\wedge \d x=\d(p^{r+1})\wedge \d\rho,\quad
\rho:= \frac{x}{(r+1)p^r}.
\]
Hence there exists an associated generating function
$J_r:=J_r(P,\rho)$ of the canonical transformation $(p,x)\mapsto
(P,X)$ such that
\[
\d J_r=p^{r+1}\d\rho+X\d P,
\]
or equivalently
\begin{equation}\label{25}
p^{r+1}=\frac{\partial J_r(P,\rho)}{\partial \rho},\quad
X=\frac{\partial J_r(P,\rho)}{\partial P}.
\end{equation}
In this way by denoting
\[
\bM_r:=\frac{\bM}{(r+1)\bL^r},
\]
we deduce
\begin{align*}
\frac{\partial}{\partial p}J_r(P(\bL,\bM),\bM_r) &=\frac{\partial
J_r }{\partial P}(P(\bL,\bM),\bM_r)\frac{\partial
P(\bL,\bM)}{\partial p}\\&+\frac{\partial J_r }{\partial
\rho}(P(\bL,\bM), \bM_r)\frac{\partial
\bM_r}{\partial p}\\
&=X(\bL,\bM)\frac{\partial P(\bL,\bM)}{\partial
p}+\bL^{r+1}\frac{\partial \bM_r}{\partial p},
\end{align*}
and by  taking into account that
\[
\bL^{r+1}\frac{\partial \bM_r}{\partial
p}=\frac{1}{r+1}\frac{\partial (\bL\bM)}{\partial
p}-\frac{\partial S}{\partial p},
\]
where $S$ is the function introduced in \eqref{7},  we deduce that
\begin{equation}\label{26}
X(\bL,\bM)=\displaystyle\frac{\displaystyle\frac{\partial}{\partial
p}\Big(
S+J_r(P(\bL,\bM),\bM_r)-\frac{1}{r+1}\bL\bM\Big)}{\displaystyle\frac{\partial}{\partial
p}P(\bL,\bM)}.
\end{equation}
This formula enables us to state
\begin{teh} In terms of the function
\begin{equation}\label{27}
\mathcal{S}_r:=S+J_r(P(\bL,\bM),\bM_r)-\frac{1}{r+1}\bL\bM,
\end{equation}
the second twistor equation $ (X(\bL,\bM))_{\leq -1}=0$ is
equivalent to the following two conditions

\noindent 1) The expansion of  $\mathcal{S}_r$ in powers of $p$
satisfies
\begin{equation}\label{28}
(\mathcal{S}_r)_{\leq -1}=0.
\end{equation}
\noindent 2) At each zero  $p_i$ of $\partial P(\bL,\bM)/ \partial
p$ it is verified that
\begin{equation}\label{29}
\frac{\partial \mathcal{S}_r}{\partial p}(p_i,\bt)=0.
\end{equation}
\end{teh}
\bigskip
\noindent Henceforth we will refer to \eqref{29} as the
\emph{hodograph equations}.
\bigskip

A natural problem is to determine generating functions
$J_r(P,\rho)$ leading to solvable twistor equations. In this
sense, an important class  arises when $P=P(p,x)$ is independent
of $x$ and has a finite-order expansion as $p\rightarrow\infty$
\[
P(p)=\sum_{n=-\infty}^N a_np^n.
\]
The corresponding generating function $J_0$ is of the form
\[
J_0(P,x)=f(P)+g(P)x,
\]
where $g(P)$ is the inverse function of $P=P(p)$. As a consequence
\begin{align*}
J_0(P(\bL,\bM),\bM)&=f(P(\bL))+\bL\bM,\\
\mathcal{S}_0&=S+f(P(\bL)).
\end{align*}
It can be shown that, provided $f(P(p))$ admits a Laurent
expansion as \newline $p\rightarrow\infty$, the twistor equations
determined by $J_0$ have a solution. Moreover, it turns out that
solving the hodograph equations for $\mathcal{S}_0$ is enough for
computing $\bL$. Let us illustrate these facts with the following
important example

\vspace{0.2 truecm}

\noindent {\bf Gelfand-Dikii reductions}

\vspace{0.2 truecm}
 If we set
\begin{equation}\label{30}
J_0(P,x)=f(P^{1/m})+P^{1/m}x,\quad
f(P^{1/m}):=\sum_{n=-\infty}^{\infty}c_nP^{n/m},
\end{equation}
for a given integer $m>1$, the associated twistor data are
\begin{equation}\label{30a}
P=p^m,\quad X=\frac{1}{mp^{m-1}}\Big(f'(p)+x\Big).
\end{equation}
Then, the first twistor equation is
\[
\bL^m=(\bL)_{\geq 0},
\]
so that
\begin{equation}\label{31}
\bL^m=p^m+q_{m-2}(\bu)p^{m-2}+\cdots+q_1(\bu)p+q_0(\bu),
\end{equation}
where the functions $q_i(\bu)$ depend on the $(m-1)$ first
coefficients $\bu:=(u_1,\ldots, u_{m-1})$ of the expansion
\eqref{1} of $\bL$. This constraint defines the $m$-th
Gelfand--Dikii reduction of the dKP hierarchy.

For example the first few reductions are
\begin{align*}
m=2,&\quad \bL^2=p^2+2u_1,\\
m=3,&\quad \bL^3=p^3+3u_1p+3u_2,\\
m=4,&\quad \bL^4=p^4+4u_1p^2+4u_2p+6u_1^2+4u_3.
\end{align*}

To determine $\bL$ we must find the $(m-1)$ unknowns  $u_i$ as
functions of $\bt$ through the second twistor equation. Thus,
according to Theorem 2 we impose
\begin{align*}
\mathcal{S}_0=&S+f(\bL)=\Big(S+f(\bL) \Big)_{\geq 0}\\
=&\sum_{n\geq 1}(t_n+c_n)(\bL^n)_{n\geq 0}+c_0.
\end{align*}
Hence, by using \eqref{31} we can express $\mathcal S_0$ as a
function of $(p,\bt,\bu)$. If we now impose the hodograph
equations \eqref{29}, we get $(m-1)$ implicit equations
\begin{equation}\label{31a}
\Big(\sum_{n\geq 1}(t_n+c_n)\frac{\partial}{\partial p}
(\bL^n)_{n\geq 0}\Big)\Big|_{p=p_i(\bu)}=0,\quad i=1,\ldots, m-1,
\end{equation}
which determine the functions $u_i(\bt)$ and, consequently, $\bL$.
Furthermore, by eliminating $p$ in \eqref{31} we can express $p$
as a function $p=p(\bL,\bt)$, which under substitution into
\[
S=\sum_{n\geq 1}t_n\bL^n-\Big(\sum_{n\geq
1}t_n\bL^n-f(\bL)\Big)_{\leq -1},
\]
leads to $\bM=\partial S/\partial \bL$. Thus, it is easy to see
that the functions $\bL$ and $\bM$ are solutions of the twistor
equations which satisfy \eqref{1} and, therefore, they solve the
dKP hierarchy Henceforth these solutions will be called
\emph{Gelfand--Dikii solutions} of the dKP hierarchy.

For instance if $m=2$ (dKdV reduction)
\[
\bL^2=p^2+2u,\quad u:=u_1,
\]
and we set $t_n=0,\; \forall n>3$, one gets the hodograph relation
\begin{equation}\label{32}
3ut+x=F(u),
\end{equation}
which solves the dKdV equation $u_t=3uu_x$. Here
\[
F(u):-=\frac{\partial}{\partial p}\sum_{n\geq
1}c_n\cdot(\bL^n)_{n\geq 0}\Big|_{p=0}.
\]
can be assumed to be an arbitrary smooth function of $u$. Some
elementary solutions provided by \eqref{32} are
\begin{equation}\label{32a}
\begin{aligned}
F(u)=&cu,\quad u=-\frac{x}{3t-c}, \\
F(u)=&cu^2,\;
u=\frac{1}{2c}\Big(3t+\sqrt{9t^2+4cx)}\Big),\\
F(u)=&cu^3,\; u=\frac{f}{2c}+\frac{2t}{f},\;
f:=\Bigg(4x+4c^2\sqrt{x^2-\frac{4t^3}{c}}\Bigg)^{\frac{1}{3}}.
\end{aligned}
\end{equation}

\subsection{The action of
additional symmetries on Gelfand--Dikii solutions }

Our aim now is to characterize solutions of the dKP hierarchy by
applying the symmetry transformations \eqref{17} to
Gelfand--Dikii solutions. Obviously we may start from solutions of
the hodograph equations \eqref{31a} and then performing the
corresponding symmetry transformation. However, in order to do it
we need to know how the coefficients $u_i$ of the expansion
\eqref{1} of $\bL$ transform under the symmetries \eqref{17},
which requires to solve a system of first-order linear partial
differential equations. We are trying instead an alternative way
consisting in determining the generating functions $J_r(P,\rho)$
for the transformed twistor data and then solving the
corresponding twistor equations according to the scheme of Theorem
3. In this alternative procedure the problem reduces to solving a
system of implicit algebraic equations.

The dKP symmetry generated by \eqref{17} acts on twistor data
according to the canonical transformation \eqref{25a}. In
particular, the twistor data \eqref{30a} for the Gelfand--Dikii
reductions transform as
\begin{equation}\label{33a}
\begin{aligned}
P(s)=&\Big(\displaystyle\frac{p}{(j_{\rho})^{\frac{1}{r+1}}}\Big)^m,\\\\
X(s)=&\frac{P^{\frac{m-1}{m}}}{m}\Big( f'(P^{1/m})+
(r+1)j\,P^{r/m}\Big).
\end{aligned}
\end{equation}
Hence, by taking into account that $j$ is a function of
$(s,\rho,h)$, it follows that
\begin{align*}
p^{r+1}&=j_{\rho}P^{\frac{r+1}{m}}=\frac{\partial}{\partial
\rho}\Big(jP^{\frac{r+1}{m}}\Big)
-\hat{h}_{\rho}j_h P^{\frac{r+1}{m}},\\
X&=\frac{\partial}{\partial P}\Big(f(P)^{1/m}+j
P^{\frac{r+1}{m}}\Big)-\hat{h}_{P}j_hP^{\frac{r+1}{m}},
\end{align*}
where
\[
\hat{h}=\hat{h}(P,\rho):=h(p(P,\rho),\rho)=\alpha(\rho)p(P,\rho)^i.
\]
By using now \eqref{26a}  we deduce
\begin{equation}\label{34a}
p^{r+1}=\frac{\partial J^{(i)}_r(P,\rho)}{\partial \rho},\quad
X=\frac{\partial J^{(i)}_r(P,\rho)}{\partial P},
\end{equation}
where
\begin{equation}\label{35a}
J^{(i)}_r(s,P,\rho):=f(P^{1/m})+j(s,\rho,\hat{h})P^{\frac{r+1}{m}}
+s\,\Big(1-\frac{i}{r+1}\Big)\hat{h}(P,\rho).
\end{equation}

Wide families of  solutions of the $(r+1)$-th truncated dKP can be
found by solving the twistor equations associated with the
generating functions \eqref{35a}. The calculations are simple but
long and require computer aid.  To illustrate the strategy for
computing these solutions let us  consider the family of
generating functions $J^{(i)}_r$ with
\begin{equation}\label{35bb}
i=r+1\geq m.
\end{equation}
The choice $i=r+1$ means that we are dealing with the orbits of
Gelfand--Dikii solutions under the action  of the symmetry
generator
\begin{equation}\label{35b}
F_{r+1}(\bL,\bM):=\alpha\Big(
\frac{\bM}{(r+1)\bL^r}\Big)\bL^{r+1}.
\end{equation}
Thus, according to \eqref{26a} the function $j$ in \eqref{35a} is
determined from $\alpha$ through the solution of the initial value
problem
\begin{equation}\label{35c}
\frac{\partial j}{\partial s}=\alpha(\rho),\quad j(0,\rho)=\rho.
\end{equation}
Hence $j$ is independent of $h$ and by setting $s$ to be a
constant,  we may take $j$ as a function of $\rho$ only.
 Therefore, the generating functions $J^{(i)}_r$ that we are considering are
\begin{equation}\label{33}
J_r(P,\rho)=f(P^{\frac{1}{m}})+j(\rho)\,P^{\frac{r+1}{m}}.
\end{equation}
Notice that
\begin{equation}\label{34}
P=\frac{p^m}{(j_{\rho})^{\frac{m}{r+1}}},
\end{equation}
so that the first twistor equation reads
\begin{equation}\label{35}
\mathbb{L}^m=(\mathbb{L}^m)_{\geq 0},
\end{equation}
where
\begin{equation}\label{36}
\mathbb{L}:=\frac{\bL}{j_{\rho}(\bM_r)},\quad
\bM_r:=\frac{\bM}{(r+1)\bL^r}.
\end{equation}
From \eqref{1} one deduces at once that the integer powers of
$\bL$ have expansions of the form
\begin{equation}\label{37}
\begin{aligned}
\bL^N=&p^N+\cdots+a_n(u_1,\ldots,u_{N-n-1})p^n
+\ldots\\+&b_n(u_1,\ldots,u_{N+n-1})\frac{1}{p^n}+\ldots,\\
\frac{1}{\bL^N}=&\frac{1}{p^N}+\ldots+c_n(u_1,\ldots,u_{n-N-1})\frac{1}{p^n}+\ldots.
\end{aligned}
\end{equation}
Furthermore, \eqref{1} implies that for any smooth function
$g=g(t)$ the composite function
 $g(\bM_r)$ can be expanded in the form
\begin{equation}\label{38}
\begin{aligned}
g(\bM_r)&=g\Big(t_{r+1}+\frac{r t_r}{r+1}\frac{1}{\bL}+\cdots+
\frac{v_n(\bt)}{r+1}\frac{1}{\bL^n}+\cdots\Big)\\
&=g(t_{r+1})+\frac{r t_r}{r+1}g'(t_{r+1})\frac{1}{p}+\cdots\\
&+d_n(\bt,u_1,\ldots,u_{n-2},v_1,\ldots,v_{n-r-1}) \frac{1}{p^n}+
\ldots.
\end{aligned}
\end{equation}
Thus, from \eqref{37}-\eqref{38} and by taking into account
\eqref{35bb}, we deduce that $\mathbb{L}$ is of the form
\begin{equation}\label{39}
\mathbb{L}=\Big(q_m(\bt,\bu)p^m+\cdots+q_1(\bt,\bu)p+q_0(\bt,\bu)\Big)^{\frac{1}{m}},
\end{equation}
where $\bu:=(u_1,\ldots,u_{m-1})$.

 Two different cases arise
\vspace{0.3 truecm}

\noindent {\bf 1. $\mathbf{r=m-1,m}$.}

\vspace{0.3 truecm}

  This is the simplest situation since from \eqref{37}-\eqref{39}
it follows at once that
\[
\mathcal{S}_r=\Big(\sum_{s=1}^r\frac{r-s+1}{r+1}t_s \bL^s+\gamma
\mathbb{L}^{m+n} +j(\bM_r)\mathbb{L}^{r+1}\Big)_{\geq 0},
\]
is a function depending of $(p,\bt)$ and
$\bu=(u_1,\ldots,u_{m-1})$. Therefore, the $(m-1)$ hodograph
equations
\begin{equation}\label{39a}
\frac{\partial \mathcal{S}_r}{\partial p}(p_i,\bt)=0,
\end{equation}
where $p_i=p_i(\bt,\bu)$ are the zeros of $\partial
\mathbb{L}^m/\partial p\;$,are enough for determining $\bu$.

\vspace{0.3 truecm}

\noindent {\bf 2.  $\mathbf{r\geq m+1}$.}

\vspace{0.3 truecm}

The function $\mathcal{S}_r=(\mathcal{S}_r)_{\geq 0}$ depends on
$(p,\bt)$ and $\widetilde{\bu}=(u_1,\ldots,u_{r-1})$, so that in
addition to the $(m-1)$ hodograph equations \eqref{39a} a set of
$(r-m)$ new equations involving $\bt$ and $\widetilde{\bu}$ are
required. These additional equations are supplied by vanishing the
coefficients of the negative powers $1/p^n \, (n=1,\ldots r-m)$ in
\[
(\mathbb{L}^m)_{\leq -1}=0.
\]

\subsection{Examples}

In the following  examples we exhibit solutions $u$ of the dKP
equation \eqref{4} depending on an arbitrary function $j=j(\rho)$.
They are
 orbits  of Gelfand--Dikii solutions $u_0$ under the action
of the symmetry generated by \eqref{35b}. Notice that according to
\eqref{35b}-\eqref{35c} we can obtain $u_0$ by setting $j=\rho$ in
the expression of $u$.

\vspace{0.3 truecm}

\noindent {\bf Examples}

\vspace{0.3 truecm}

\vspace{0.3 truecm}

\noindent {\bf 1}. For
\[
r=m=2,\quad f(P^{\frac{1}{2}}):=\gamma P^{\frac{7}{2}},
\]
the generating function \eqref{33} becomes
\begin{equation}\label{43}
J_2(P,\rho)=\gamma P^{\frac{7}{2}}+j(\rho)P^{\frac{3}{2}},\quad
\rho:=\frac{x}{3p^2},
\end{equation}
and $\mathbb{L}^2$ takes the form
\begin{align}\label{44}
\nonumber \mathbb{L}^2=&(\mathbb{L}^2)_{\geq
0}=\frac{p^2}{(j'(t))^{2/3}}-
\frac{4}{9}\frac{yj''(t)}{(j'(t))^{5/3}}p+\frac{2u_1}{(j'(t))^{2/3}}\\\\
\nonumber -&\frac{2}{9}\frac{xj''(t)}{(j'(t))^{5/3}}-\frac{4}{27}
\frac{y^2j'''(t)}{(j'(t))^{5/3}}+\frac{20}{81}\frac{y^2(j''(t))^2}{(j'(t))^{8/3}}.
\end{align}
Hence $ \partial \mathbb{L}^2 / \partial p$ has a unique zero
given by
\[
p_1=\frac{2}{9}y\frac{j''(t)}{j'(t)}.
\]
 Moreover the expression of
\[
\mathcal{S}_2=\Big(\frac{1}{3}y\bL^2+\frac{2}{3}x\bL+\gamma
\mathbb{L}^7+ j(\bM_2)\mathbb{L}^3\Big)_{\geq 0}.
\]
as a function of $p$ can be computed by using \eqref{44} and the
expansion
\begin{align*}
j(\bM_2)&=j(t)+\frac{2}{3}yj'(t)\frac{1}{p}+\Big(\frac{x}{3}j'(t)+
\frac{2}{9}y^2j''(t)\Big)\frac{1}{p^2}\\
&+\Big(-\frac{2}{3}yj'(t)u_1+\frac{4}{81}y^3j'''(t)+\frac{2}{9}xyj''(t)\Big)
\frac{1}{p^3}+\mathcal{O}\Big(\frac{1}{p^4}\Big).
\end{align*}
In this way the hodograph equation $(\partial \mathcal{S}_2
/\partial p)|_{p=p_1}=0$ turns out
 to be an equation for $u=u_1$, which yields
the following solution of the dKP equation

\begin{align}\label{44}
\nonumber u&=\frac{F}{105\gamma}-\frac{6j(t)j'(t)^{4/3}}{F}\\
&+\frac{9j'(t)j''(t)x+6j'(t)j'''(t)y^2-8j''(t)^2y^2}{81(j'(t))^2},
\end{align}
where
\begin{align*}
F:&=\gamma^{2/3}\Big(-7350j'(t)^{4/3}j''(t)y^2-33075j'(t)^{7/3}x+
105\sqrt{35\,G}\Big)^{1/3}
,\\
G:&=\frac{1}{\gamma}\Big(648j(t)^3j'(t)^4+140\gamma
j'(t)^{8/3}j''(t)^2y^4+1260\gamma j'(t)^{11/3}j''(t)xy^2\\
&+2835\gamma j'(t)^{14/3}x^2\Big).
\end{align*}

\vspace{0.3 truecm}

\noindent {\bf 2}. By setting
\[
r=m=3,\quad f(P^{1/3}):=\gamma P^{7/3},
\]
in \eqref{33} one finds that the  first two coefficients of the
expansion \eqref{1} of $\bL$ are given by
\begin{equation}\label{45}
 u=u_1=-\frac {1}{1024\,j_1^2}\Big( 90\,j_2^2 t^{2}-72\,
j_1j_3t^2 -128\,j_1 j_2 y+Z^2\Big),
\end{equation}
\begin{equation}\label{46}
u_2=\frac{-21\,\gamma j_1^8j_2t\, Z^4+F\,Z^2+8388608\,
j_1^{\frac{59}{4}}y+2359296\,j_1^{\frac{55}{4}}j_2t^2}
{114688\,\gamma j_1^{11}\,Z}.
\end{equation}
where
\begin{align*}
j_i:&=\frac{\partial^i j}{\partial \rho^i}(t_4),\quad i\geq 0,
\\
F:=&-16384\,j_0j_1^{\frac{47}{4}}+7168\,\gamma j_1^{10}j_2
 x-13440\,\gamma
j_1^{9}j_2^2 ty+5670\,\gamma j_1^{8}j_2^3 t^3\\
+&2016\, \gamma j_1^{10}j_4 t^3-7560\,\gamma j_1^{9}j_2 j_3
t^3+10752\, \gamma j_1^{10}j_3 ty,
\end{align*}
and $Z=Z(x,y,t,t_4)$ is a root of the equation
\begin{align*}
 &49\,j_1^{30}\gamma^2 Z^{10} +\Big( 5637144576\,
 \gamma j_1^{\frac {151}{4}}x
 +2113929216\,\gamma j_1^{\frac {147}{4}}j_2\,ty\\
&-1610612736\, j_0^2 j_1^{\frac {75}{2}}+396361728\,\gamma
j_1^{\frac {147}{4}}
j_3t^3\\
&-297271296\,\gamma\, j_1^{\frac {143}{4}}j_2^{2}t^3 \Big)
 Z^4+422212465065984
\, j_1^{\frac {87}{2}}y^2
\\&+33397665693696\,j_1^{\frac {83}{2}}
j_2^{2}t^4 +237494511599616\,j_1^{\frac {85}{2}}j_2t^2y=0.
\end{align*}

\noindent {\bf Acknowledgements}

The authors are grateful to Prof. F. Guil for showing them the
relevance of generating functions for canonical transformations of
twistor data to get solutions of the dKP hierarchy.

\end{document}